\documentclass[12pt]{article}

\usepackage{epsfig}

\def\square{\kern1pt\vbox{\hrule height 1.2pt
\hbox{\vrule width 1.2pt\hskip 3pt
\vbox{\vskip 6pt}\hskip 3pt\vrule width 0.6pt}
\hrule height 0.6pt}\kern1pt}
\def\ltwid{\mathrel{\raise.3ex\hbox{$<$\kern-.75em\lower1ex\hbox{$\sim$}}}}
\def\gtwid{\mathrel{\raise.3ex\hbox{$>$\kern-.75em\lower1ex\hbox{$\sim$}}}}

\begin{document}

\begin{titlepage}
\begin{flushright}
CCTP-2014-08 \\ UFIFT-QG-14-03
\end{flushright}

\vspace{0.5cm}

\begin{center}
\bf{Classical Gravitational Back-Reaction}
\end{center}

\vspace{0.3cm}

\begin{center}
N. C. Tsamis$^{\dagger}$
\end{center}
\begin{center}
\it{Institute of Theoretical \& Computational Physics, and \\
Department of Physics, University of Crete \\
GR-710 03 Heraklion, HELLAS.}
\end{center}

\vspace{0.2cm}

\begin{center}
R. P. Woodard$^{\ast}$
\end{center}
\begin{center}
\it{Department of Physics, University of Florida \\
Gainesville, FL 32611, UNITED STATES.}
\end{center}

\vspace{0.3cm}

\begin{center}
ABSTRACT
\end{center}
\hspace{0.3cm} 
The quantum gravitational back-reaction on inflation is based on 
the self-gravitation of infrared gravitons which are ripped out 
of the vacuum during inflation. The only quantum part of this process 
is the creation of gravitons; after they have emerged from the vacuum 
their behaviour is essentially classical. To test the thesis that a 
sufficiently dense ensemble of classical gravitons can hold the universe 
together in pure gravity with a positive cosmological constant, we
compute the initial value and first time derivative of an invariant 
measure of the expansion rate for arbitrary classical initial value 
data. Our result is that the self-gravitation from the kinetic energy 
of an initial ensemble of gravitons can indeed slow expansion enough 
to hold the universe together.

\vspace{0.3cm}

\begin{flushleft}
\end{flushleft}

\vspace{0.1cm}

\begin{flushleft}
$^{\dagger}$ {\it e-mail:} tsamis@physics.uoc.gr \\
$^{\ast}$ {\it e-mail:} woodard@phys.ufl.edu
\end{flushleft}

\end{titlepage}

\section{Introduction}

Gravitation plays the dominant role in shaping cosmological 
evolution. Moreover, a wide variety of observational evidence
points to the very early universe having experienced a phase 
of accelerated expansion, or inflation \cite{infl}. During 
inflation, quantum physics implies the production of real 
particles out of the vacuum as long as they are effectively 
massless, possess classically non-conformally invariant free
Lagrangians, and have adequately large wavelength. The carrier 
of the gravitational force, the graviton, is such a particle 
and inflationary evolution eventually will produce a dense 
ensemble of infrared gravitons \cite{Grishchuk}.

Gravitation couples to any stress-energy source and the 
aforementioned quantum induced source of gravitons is no 
exception. It becomes important, therefore, to study the 
gravitational response to its presence. Being a universally
attractive force, gravity has the potential to alter the
inflationary expansion rate and decrease it. This has already
been suggested \cite{TW1} but the supporting perturbative 
analysis eventually becomes unreliable; the self-gravitation 
of the infrared gravitons ripped out of the vacuum very slowly 
but cumulatively increases until perturbation theory breaks 
down.

That said, the question arises whether we can make any 
quantitative non-perturbative statements. With this in mind, 
we note that our physical problem can be stated as the 
{\it classical} gravitational back-reaction to a {\it quantum} 
induced graviton source; only the particle creation out of 
the vacuum is a quantum effect. Detailed knowledge of this 
quantum source would allow similar knowledge of the response 
to its presence, the latter being determined by the field 
equations of gravity. However, the non-linearity of the 
theory is a formidable hindrance both for the description 
of the graviton source and the response to it. 

Nonetheless, it is possible to obtain non-perturbatively some 
measure of the back-reaction on an initial value surface (IVS) 
for arbitrary initial value data (IVD). In the real situation
the initial value surface would coincide with the beginning
of the inflationary era and after considerable time evolution 
the quantum induced graviton source would slowly but steadily
become significant. Even if we lack an analytical form for the
source, it {\it must} correspond to some IVD. Even if full 
time evolution is beyond our means, we {\it can} compute the 
expansion rate and its first time derivative on the IVS. 

A physical measure of the back-reaction can be provided by 
an observable which invariantly determines the expansion rate 
\cite{TW2} and which we review in Section 2. The computation 
of its initial value and first time derivative for any classical 
initial value data are presented in Section 3. Our concluding 
remarks comprise Section 4.

\section{The Expansion Rate}

In the presence of a cosmological constant $\Lambda$ the 
gravitational field equations are:
\footnote{Hellenic indices take on spacetime values 
while Latin indices take on space values. Our metric
tensor $g_{\mu\nu}$ has spacelike signature
$( - \, + \, + \, +)$ and our curvature tensor equals
$R^{\alpha}_{~ \beta \mu \nu} \equiv 
\Gamma^{\alpha}_{~ \nu \beta , \mu} +
\Gamma^{\alpha}_{~ \mu \rho} \, 
\Gamma^{\rho}_{~ \nu \beta} -
(\mu \leftrightarrow \nu)$.}
\begin{equation}
R_{\mu\nu} - \frac12 \, g_{\mu\nu} \, R + \Lambda \, g_{\mu\nu} 
\, = \, 0
\;\; . \label{eom}
\end{equation}
The standard local definition of the expansion rate 
${\cal H}$ \cite{HawkEll}:
\begin{equation}
{\cal H}(t, {\bf x})
\, = \, 
\frac13 \, D^{\mu} u_{\mu}(t, {\bf x})
\;\; , \label{H}
\end{equation}
is in terms of the covariant derivative $D_{\mu}$ of 
a timelike 4-velocity field $u_{\mu}$:
\begin{equation}
g^{\mu\nu}(x) \, u_{\mu}(x) \, u_{\nu}(x) 
\, = \, - 1
\;\; . \label{u}
\end{equation}

An appropriate 4-velocity field can be constructed from 
a scalar functional $\Phi$ of the metric satisfying, for 
all $x$, the dynamical equation:
\footnote{The construction that follows has been described
in detail in \cite{TW2}. Further approaches to invariant 
expansion observables can be found in \cite{BG,FMVV,MVB,MV}.}
\begin{equation}
\square \Phi[g](x) \, = \,
\frac{1}{\sqrt{-g}} \,
\partial_{\mu} [\sqrt{-g} \, g^{\mu\nu} \, \partial_{\nu} \Phi]
\, = \,
3 H
\;\; , \label{Phi}
\end{equation}
where $H$ is the Hubble parameter ($\Lambda = 3 H^2$). 
On the initial value surface the scalar $\Phi$ satisfies:
\begin{equation}
\Phi(t_I, {\bf x}) \Big\vert_{\rm IVS} \, = \, 0
\quad , \quad
-g^{\alpha\beta}(t_I, {\bf x}) \; 
\partial_{\alpha} \Phi(t_I, {\bf x}) \;
\partial_{\beta} \Phi(t_I, {\bf x}) 
\Big\vert_{\rm IVS} \, = \, 1
\;\; . \label{IVD}
\end{equation}
The resulting 4-velocity field $V_{\mu}$ equals:
\begin{equation}
V_{\mu} [g](x) \, \equiv \,
+ \frac{\partial_{\mu} \Phi[g](x)}
{\sqrt{-g^{\alpha\beta}(x) \; \partial_{\alpha} \Phi[g](x) \;
\partial_{\beta} \Phi[g](x)}}
\;\; , \label{V}
\end{equation}
and the expansion variable according to (\ref{H}) is: 
\begin{equation}
{\cal H}[g](x) \, = \,
\frac13 \, D^{\mu} V_{\mu}[g](x)
\, = \,
\frac13 \, \frac{1}{\sqrt{-g}} \,
\partial_{\mu} [\sqrt{-g} \, g^{\mu\nu} \, V_{\nu}]
\;\; . \label{H2}
\end{equation}
We can invariantly fix the observation time by specifying
the surfaces of simultaneity as follows:
\begin{equation}
\Phi[g](\vartheta [g](x), {\bf x}) \, = \,
\Phi_{\rm dS} (t)
\;\; , \label{time}
\end{equation}
where $\Phi_{\rm dS} (t)$ is the scalar $\Phi$ in de Sitter 
spacetime. This requirement determines the functional 
$\vartheta [g](x)$ or, equivalently, the observation time.

Our observable ${\rm H}$ -- which physically represents 
the expansion rate of spacetime -- is given by :
\begin{equation}
{\rm H}[g](x) \, \equiv \,
{\cal H}[g](\vartheta [g](x), {\bf x})
\;\; . \label{H3}
\end{equation}
Under general coordinate transformations which preserve the 
initial value surface, the variable just constructed transform 
thusly:
\begin{equation}
{\cal H}[g'] (x) \, = \, {\cal H}[g] (x'^{\, -1}(x))
\quad , \quad
{\rm H}[g'] (t, {\bf x}) \, = \, 
{\rm H}[g] (t, x'^{\, -1}(t, {\bf x}))
\;\; . \label{transf}
\end{equation}

\section{The Classical Computation on the Initial Value Surface}

We now turn to the main results of this study, the calculation
of the value and first time derivative of the expansion rate
observable on the IVS. Because $\Phi \vert_{\rm IVS} = 0$, the 
invariant observation time condition (\ref{time}) is automatically 
satisfied on the IVS and need not concern us. Consequently, it 
suffices to consider the expansion rate as provided by ${\cal H}$. 
\\ [-9pt] 

$\bullet \,$ {\it The $3+1$ decomposition.} \\
The nature of our problem suggests that we employ a coordinate
system that separates space and time.
\footnote{The pioneering work on the subject by Arnowitt, 
Deser and Misner (ADM) can be found in \cite{ADM}; see also 
\cite{Cook,SY}.}
The $3+1$ decomposition of the line element is:
\begin{eqnarray}
ds^2 &\!\! = \!\!&
- g_{00} dt^2 + 2 g_{0i} \, dt dx^i + g_{ij} \, dx^i dx^j
\nonumber \\
&\!\! = \!\!&
- N^2 dt^2 + \gamma_{ij} \, (dx^i + N^i dt) (dx^j + N^j dt)
\;\; , \label{ds^2}
\end{eqnarray}
with $N$ the lapse, $N^i$ the shift vector and $\gamma_{ij}$ 
the spatial metric. It follows that the elements of the 
the spacetime metric $g_{\mu\nu}$ are:
\begin{equation}
g_{00} = - N^2 + N_i N^i
\quad , \quad
g_{0i} = N_i
\quad , \quad
g_{ij} = \gamma_{ij}
\;\; , \label{g_mn}
\end{equation}
while those of its inverse metric $g^{\mu\nu}$ are:
\begin{equation}
g^{00} = - \frac{1}{N^2}
\quad , \quad 
g^{0i} = \frac{N^i}{N^2}
\quad , \quad
g^{ij} = \gamma^{ij} - \frac{N^i N^j}{N^2}
\;\; . \label{g^mn}
\end{equation}



The relevant Christoffel connections are:
\footnote{Henceforth, a bar over a symbol indicates it is
purely spatial, a comma indicates a derivative with respect
to the spatial metric $\gamma_{ij}$, and a semicolon a 
covariant derivative.}
\begin{eqnarray}
\Gamma^{0}_{~ 00} \!\!& = &\!\!
\frac{N_{, 0}}{N} + \frac{N^i N_{, i}}{N}
- \frac{N^i N^j N_{i ; j}}{N^2}
\;\; , \label{000} \\
\Gamma^{0}_{~ 0i} \!\!& = &\!\!
\frac{N_{, i}}{N} - \frac{N^j K_{ij}}{N}
\;\; , \label{00i} \\
\Gamma^{0}_{~ ij} \!\!& = &\!\!
- \frac{K_{ij}}{N}
\;\; , \label{0ij} \\
\Gamma^{i}_{~ 00} \!\!& = &\!\!
- \frac{N^i}{N} \Big[ N_{, 0} + N^j N_{, j} - N^j N^k K_{jk} \Big] 
+ N^i_{, 0} - N N^{, i} - 2 N N^j K^i_j 
\nonumber \\
& \mbox{} &
+ N^j N^i_{; j}
\;\; , \label{i00} \\ 
\Gamma^{i}_{~ j0} \!\!& = &\!\!
- \frac{N^i N_{, j}}{N} + N^i_{; j}
- \Big( \gamma^{ik} - \frac{N^i N^j}{N^2} \Big) N K_{kj}
\;\; , \label{ij0} \\ 
\Gamma^{i}_{~ jk} \!\!& = &\!\!
\frac{N^i K_{jk}}{N} + {\bar{\Gamma}}^i_{~ jk} 
\;\; , \label{ijk}
\end{eqnarray}
where $K_{ij}$ is the extrinsic curvature.

The gravitational field equations (\ref{eom}) can be separated
into evolution equations and constraints. The former are:
\begin{eqnarray} 
\partial_0 \gamma_{ij} \!\!& = &\!\!
-2N K_{ij} + {\bar D}_i N_j + {\bar D}_j N_i
\;\; , \label{eom1} \\
\partial_0 K_{ij} \!\!& = &\!\!
- {\bar D}_i {\bar D}_j N + N^k {\bar D}_k K_{ij} 
+ K_{ik} {\bar D}_j N^k + K_{jk} {\bar D}_i N^k
\nonumber \\
& \mbox{} & 
+ N \Big[ {\bar R}_{ij} - 2 K_{ik} K^k_j + K K_{ij} 
- 3H^2 \gamma_{ij} \Big]
\;\; , \label{eom2}
\end{eqnarray}
while the latter take the form:
\begin{eqnarray}
{\bar R} + K^2 - K_{ij} K^{ij} \!\!& = &\!\! 6 H^2
\;\; , \label{constr1} \\
{\bar D}_j ( K^{ij} - \gamma^{ij} K ) \!\!& = &\!\! 0
\;\; , \label{constr2} 
\end{eqnarray}
where $K \equiv \gamma^{ij} K_{ij}$ is the trace of the 
extrinsic curvature and ${\bar D}_i$ is the spatial
covariant derivative with respect to $\gamma_{ij}$.

In this decomposition, there are $12=6+6$ canonical degrees 
of freedom that $\gamma_{ij}$ and $K_{ij}$ contain. Of these, 
only $4=2+2$ are dynamical and correspond to the two 
polarization states of the graviton; the other are the $4$ 
constrained degrees of freedom from the $1+3$ constraint 
equations (\ref{constr1}-\ref{constr2}), and the $4$ gauge 
degrees of freedom from the initial coordinate system choices.
\\ [-9pt]

$\bullet \,$ {\it The elements of the observable on the IVS.} \\
The equation of motion (\ref{Phi}) of the scalar $\Phi$ is:
\begin{eqnarray}
3 H = 
\square \Phi = 
g^{\mu\nu} D_{\mu} D_{\nu} \Phi 
&\!\! = \!\!&
g^{00} D_0 D_0 \Phi +
2 g^{0i} D_0 D_i \Phi +
g^{ij} D_i D_j \Phi 
\qquad \label{eomPhi1} \\
&\!\! = \!\!&
g^{\mu\nu} ( \Phi_{, \mu\nu} - \Gamma^{\rho}_{\mu\nu} \Phi_{, \rho} )
\;\; . \label{eomPhi2} 
\end{eqnarray}
The initial value conditions (\ref{IVD}) on the scalar $\Phi$ can
be conveniently written as:
\begin{equation}
\Phi \, \Big\vert_{\rm IVS} \, = \, 0
\quad , \quad
\Phi_{, \mu} \, \Big\vert_{\rm IVS} \, = \,
- N \, \delta^0_{\mu}
\;\; . \label{PhiIVS}
\end{equation}
We shall also need the following initial covariant derivatives
of $\Phi$:
\begin{eqnarray}
D_0 D_0 \Phi \, \Big\vert_{\rm IVS} \!\!& = &\!\!
\Big[ \Phi_{, 00} - \Gamma^{\rho}_{~ 00} \Phi_{, \rho} 
\Big] \, \Big\vert_{\rm IVS}
\, = \,
- N^2 ( 3H + K ) - N^i N^j K_{ij}
\;\; , \label{DoDoPhi} \\
D_0 D_i \Phi \, \Big\vert_{\rm IVS} \!\!& = &\!\!
\Big[ \Phi_{, 0i} - \Gamma^{\rho}_{~ 0i} \Phi_{, \rho} 
\Big] \, \Big\vert_{\rm IVS}
\, = \,
- N^j K_{ji}
\;\; , \label{DoDiPhi} \\
D_i D_j \Phi \, \Big\vert_{\rm IVS} \!\!& = &\!\!
\Big[ \Phi_{, ij} - \Gamma^{\rho}_{~ ij} \Phi_{, \rho} 
\Big] \, \Big\vert_{\rm IVS}
\, = \,
- K_{ij}
\;\; , \label{DiDjPhi} \\
D_0 D_i D_j \Phi \, \Big\vert_{\rm IVS} \!\!& = &\!\!
\Big[ (D_i D_j \Phi)_{, 0} - \Gamma^{\rho}_{~ 0i} \, D_{\rho} D_j \Phi
- \Gamma^{\rho}_{~ 0j} \, D_i D_{\rho} \Phi
\Big] \, \Big\vert_{\rm IVS}
\label{DoDiDjPhia} \\
\!\!& = & \!\!
+ 3 N H^2 \gamma_{ij} - 3 H N K_{ij} - 2 N K K_{ij}
- \frac{1}{N} K_{ij} N^k N^l K_{kl} 
\qquad \nonumber \\
& \mbox{} &
+ \frac{1}{N^2} K_{ij} N^k N^l N_{k ; l} - N^k K_{ij ; k}
- N {\bar R}_{ij}
\;\; . \label{DoDiDjPhib}
\end{eqnarray}

In view of (\ref{PhiIVS}), the 4-velocity field (\ref{V})
becomes:
\begin{equation}
V_{\mu} \,\Big\vert_{\rm IVS} \,  = \,
- N \, \delta^0_{\mu}
\quad , \quad
V^{\mu} \,\Big\vert_{\rm IVS} \,  = \,
+ \frac{1}{N} \, \delta^{\mu}_0
\;\; . \label{VIVS}
\end{equation}

$\bullet \,$ {\it The observable on the IVS.} \\
The general form of the local expansion rate is given by
(\ref{H2}):
\begin{eqnarray}
{\cal H} \!\!& = & \!\!
\frac13 D_{\mu} V^{\mu} \, = \,
\frac13 \; \frac{1}{\sqrt{-g}} \;
\partial_{\mu} \left(
\frac{\sqrt{-g} g^{\mu\nu} \Phi_{,\nu}}
{\sqrt{-g^{\alpha\beta} \Phi_{,\alpha} \Phi_{,\beta}}} \right)
\label{H2a} \\
\!\! & = &\!\!
\frac13 \;
\frac{\square \Phi} 
{\sqrt{-g^{\alpha\beta} \Phi_{,\alpha} \Phi_{,\beta}}}
\; + \; 
\frac{g^{\mu\nu} \Phi_{,\mu} \; g^{\rho\sigma} \Phi_{,\rho} \; 
D_{\nu} D_{\sigma} \Phi}
{3 \, ( -g^{\alpha\beta} \Phi_{,\alpha} \Phi_{,\beta} )^{\frac32}}
\;\; . \label{H2b}
\end{eqnarray}
When restricting to the initial value surface we sequentially 
obtain:
\begin{eqnarray}
{\cal H} \, \Big\vert_{\rm IVS} \!\!& = &\!\!
\frac13 \, \Big( 3H + 
g^{\mu\nu} \Phi_{,\mu} \; g^{\rho\sigma} \Phi_{,\rho} \; 
D_{\nu} D_{\sigma} \Phi \Big) \Big\vert_{\rm IVS}
\label{H2IVSa} \\
\!\!& = &\!\!
\frac13 \, \Big( 3H + 
N^2 g^{0 \nu} g^{0 \sigma} \, 
D_{\nu} D_{\sigma} \Phi \Big) \Big\vert_{\rm IVS}
\label{H2IVSb} \\
& \mbox{} &
\hspace{-1.6cm}
= \,
\frac13 \, \Big\{ 3H + 
N^2 \Big[ 
g^{00} g^{00} \, D_0 D_0 \Phi + 
2 g^{00} g^{0i} \, D_0 D_i \Phi +
g^{0i} g^{0j} \, D_i D_j \Phi 
\Big] \Big\} \Big\vert_{\rm IVS}
\qquad \label{H2IVSc} \\
\!\!& = &\!\!
\frac13 \, \Big\{ 3H + 
N^2 \Big[ 
- \! N^{-2} \Big( 3H - g^{ij} D_i D_j \Phi \Big) +
g^{0i} g^{0j} \, D_i D_j \Phi 
\Big] \Big\} \Big\vert_{\rm IVS}
\label{H2IVSd} \\
\!\!& = &\!\!
\frac13 \, \gamma^{ij} D_i D_j \Phi \, \Big\vert_{\rm IVS}
\, = \, 
- \frac13 \, \gamma^{ij} K_{ij} \, \Big\vert_{\rm IVS}
\,= \, 
- \frac13 \, K \, \Big\vert_{\rm IVS}
\label{H2IVSe} 
\end{eqnarray}
where -- besides the form of the metric (\ref{g^mn}) and the 
double covariant derivative (\ref{DiDjPhi}) of $\Phi$ -- we 
have used the equation of motion (\ref{eomPhi1}). Since $K$ 
is a pure gauge degree of freedom we conclude that ${\cal H}$ 
can take {\it any} initial value of our choice. Therefore,
we can make it vanish on the IVS by the gauge choice $K = 0$
and then ask whether it will stay zero under time evolution.
\\ [-9pt] 

$\bullet \,$ {\it The first time derivative of the observable 
on the IVS.} \\
In order to investigate the behaviour of the observable under
infinitesimal time evolution, we consider its first derivative:
\begin{eqnarray}
D_{\mu} {\cal H} \!\!& = &\!\!
\frac{H g^{\kappa\lambda} \Phi_{,\kappa} \, D_{\mu} D_{\lambda} \Phi}
{(- g^{\alpha\beta} \Phi_{,\alpha} \Phi_{,\beta})^{\frac32}}
+
\frac{H g^{\kappa\lambda} \Phi_{,\kappa} \, 
g^{\rho\sigma} \Phi_{,\rho} \, (D_{\lambda} D_{\sigma} \Phi) \,
g^{\gamma\delta} \Phi_{,\gamma} \, D_{\mu} D_{\delta} \Phi}
{(- g^{\alpha\beta} \Phi_{,\alpha} \Phi_{,\beta})^{\frac52}}
\nonumber \\
& \mbox{} &
+ \, \frac{\frac23 g^{\kappa\lambda} \Phi_{,\kappa} \,
g^{\rho\sigma} (D_{\mu} D_{\rho} \Phi) D_{\lambda} D_{\sigma} \Phi
+ \frac13 g^{\kappa\lambda} \Phi_{,\kappa} \,
g^{\rho\sigma} \Phi_{,\rho} \, D_{\mu} D_{\lambda} D_{\sigma} \Phi}
{(- g^{\alpha\beta} \Phi_{,\alpha} \Phi_{,\beta})^{\frac32}}
\;\; , \qquad \label{DPhi}
\end{eqnarray}
which on the initial value surface the reduces to:
\begin{eqnarray}
D_{\mu} {\cal H} \, \Big\vert_{\rm IVS} 
\!\!& = &\!\! 
H g^{0 \lambda} (-N D_{\mu} D_{\lambda} \Phi)
+
N^2 g^{0 \lambda} g^{0 \sigma} D_{\lambda} D_{\sigma}
(-N g^{0 \delta} D_{\mu} D_{\delta} \Phi)
\nonumber \\
& \mbox{} &
- \, \frac23 N g^{0 \lambda} g^{\rho\sigma} 
(D_{\mu} D_{\rho} \Phi) D_{\lambda} D_{\sigma} \Phi
+
\frac13 N^2 g^{0 \lambda} g^{0 \sigma} 
D_{\mu} D_{\lambda} D_{\sigma} \Phi
\;\; . \qquad \label{DPhiIVS} 
\end{eqnarray}
We shall be interested in the $\mu = 0$ component of 
(\ref{DPhiIVS}) that we write schematically as the sum 
of four terms, respectively corresponding to the four 
terms of (\ref{DPhiIVS}):
\begin{equation}
\partial_0 {\cal H} \, \Big\vert_{\rm IVS}
\, \equiv \,
I_1 + I_2 + I_3 + I_4
\;\; . \label{DoPhiIVS1}
\end{equation}
For their reduction, the derivative of the $\Phi$ equation 
of motion (\ref{eomPhi1}):
\begin{equation}
g^{00} D_{\mu} D_0 D_0 \Phi +
2 g^{0i} D_{\mu} D_0 D_i \Phi +
g^{ij} D_{\mu} D_i D_j \Phi \, = \, 0
\;\; , \label{DeomPhi}
\end{equation}
has been useful:
\begin{eqnarray}
I_1 &\!\! \equiv \!\!&
H g^{0 \lambda} (-N D_0 D_{\lambda} \Phi)
\nonumber \\
&\!\! = \!\!&
- N H \Big[ 3H - g^{0i} D_0 D_i \Phi - g^{ij} D_i D_j \Phi \Big]
\,\, , \label{I1} \\
I_2 &\!\! \equiv \!\!&
N^2 g^{0\lambda} g^{0\sigma} D_{\lambda} D_{\sigma}
(-N g^{0\delta} D_0 D_{\delta} \Phi)
\nonumber \\
&\!\! = \!\!&
- N^3 \Big\{ g^{00} \, [ 3H - g^{ij} D_i D_j \Phi ]
+ g^{0i} g^{0j} D_i D_j \Phi \Big\} 
\nonumber \\
& \mbox{} & 
\times 
\Big[ 3H - g^{0k} D_0 D_k \Phi - g^{kl} D_k D_l \Phi \Big]
\;\; , \label{I2} \\
I_3 &\!\! \equiv \!\!&
- \frac23 N g^{0 \lambda} g^{\rho\sigma} (D_0 D_{\rho} \Phi) 
D_{\lambda} D_{\sigma} \Phi
\nonumber \\
\!\!& = &\!\!
- \frac23 N \Big\{ 
9H^2 + 
3H \Big[ -2 g^{ij} D_i D_j \Phi - g^{0i} D_0 D_i \Phi
+ \frac{g^{0i} g^{0j}}{g^{00}} D_i D_j \Phi \Big]
\nonumber \\
& \mbox{} &
+ \Big[ - g^{0i} g^{0j} + g^{00} g^{ij} \Big]
(D_0 D_i \Phi) D_0 D_j \Phi
\nonumber \\
& \mbox{} &
+ \Big[ g^{ij} g^{kl} - \frac{g^{0i} g^{0j}}{g^{00}} g^{kl} \Big]
(D_i D_j \Phi) D_k D_l \Phi
\nonumber \\
& \mbox{} &
+ \Big[ g^{0i} g^{jk} - 2 g^{0i} \frac{g^{0j} g^{0k}}{g^{00}} 
+ g^{0j} g^{ik} \Big]
(D_0 D_i \Phi) D_j D_k \Phi \Big\}
\;\; , \label{I3} \\
I_4 &\!\! \equiv \!\!&
\frac13 N^2 g^{0 \lambda} g^{0 \sigma} 
D_0 D_{\lambda} D_{\sigma} \Phi
\nonumber \\
&\!\! = \!\!&
\frac13 N^2 \Big[ - g^{00} g^{ij} + g^{0i} g^{0j} \Big]
D_0 D_i D_j \Phi
\;\; . \label{I4}
\end{eqnarray}

Grouping together the terms from expansions (\ref{I1}-\ref{I4})
according to their $H$ content we notice that:
\\ [5pt]
{\it (i)} the terms proportional to $H^2$ cancel when added up,
\\ [3pt]
{\it (ii)} the terms proportional to $H$ add up to,
\begin{equation}
J_1 \, \equiv \,
N H \Big[ - g^{ij} - N^2 g^{0i} g^{0j} \Big] D_i D_j \Phi 
\, = \,
- N H \gamma^{ij} D_i D_j \Phi
\, = \,
+ N H K
\;\; , \label{J1}
\end{equation}
{\it (iii)} the terms without $H$ dependence -- and organized 
according to their covariant derivatives structure -- are,
\begin{eqnarray}
J_2 &\!\! \equiv \!\!&
N (D_0 D_i \Phi) D_0 D_j \Phi 
\times
\frac{2}{3N^2} \Big[ g^{ij} + N^2 g^{0i} g^{0j} \Big]
\label{J2a} \\
&\!\! = \!\!&
+ \frac{2}{3N} \gamma^{ij}
(D_0 D_i \Phi) D_0 D_j \Phi \, = \,
+ \frac{2}{3N} N^i N^j K^k_i K_{kj}
\;\; , \label{J2b} \\ 
J_{3a} &\!\! \equiv \!\!&
N (D_0 D_i \Phi) D_j D_k \Phi 
\times
\frac13 \, g^{0i} \Big[ g^{jk} + N^2 g^{0j} g^{0k} \Big]
\label{J3aa} \\
&\!\! = \!\!&
+ \frac{1}{3N} N^i \gamma^{jk}
(D_0 D_i \Phi) D_j D_k \Phi \, = \,
+ \frac{1}{3N} K N^i N^j K_{ij}
\;\; , \label{J3ab} \\
J_{3b} &\!\! \equiv \!\!&
N (D_0 D_i \Phi) D_j D_k \Phi 
\times
\frac23 \, g^{0j} \Big[ - g^{ik} - N^2 g^{0i} g^{0k} \Big]
\label{J3ba} \\
&\!\! = \!\!&
- \frac{2}{3N} N^j \gamma^{ik}
(D_0 D_i \Phi) D_j D_k \Phi \, = \,
- \frac{2}{3N} N^i N^j K^k_i K_{kj}
\;\; , \label{J3bb} \\
J_4 &\!\! \equiv \!\!&
N (D_i D_j \Phi) D_k D_l \Phi 
\times
\frac13 \, g^{ij} \Big[ g^{kl} + N^2 g^{0k} g^{0l} \Big]
\label{J4a} \\
&\!\! = \!\!&
+ \frac{1}{3} N \Big[ \gamma^{ij} - 
\frac{N^i N^j}{N^2} \Big] \gamma^{kl}
(D_i D_j \Phi) D_k D_l \Phi 
\nonumber \\
&\!\! = \!\!&
+ \frac{1}{3} N K^2 - \frac{1}{3N} K N^i N^j K_{ij}
\;\; , \label{J4b} \\ 
J_5 &\!\! \equiv \!\!&
D_0 D_i D_j \Phi 
\times
\frac13 \, \Big[ g^{ij} + N^2 g^{0i} g^{0j} \Big]
\label{J5a} \\
&\!\! = \!\!&
+ \frac13 \gamma^{ij}
D_0 D_i D_j \Phi 
\, = \,
+ 3N H^2 - \frac13 {\bar R} - N H K - \frac23 N K^2
\nonumber \\
& \mbox{} &
- \frac13 N^i K_{,i} - \frac{1}{3N} K N^i N^j K_{ij}
+\frac{1}{3N^2} K N^i N^j N_{i ; j}
\;\; . \label{J5b}
\end{eqnarray}

The final result is the sum of the above {\it (ii)} 
and {\it (iii)} terms:
\begin{eqnarray}
\partial_0 {\cal H} \, \Big\vert_{\rm IVS} 
&\!\! = \!\!&
J_1 + J_2 + J_{3a} + J_{3b} + J_4 + J_5
\label{DoPhiIVS2} \\
&\!\! = \!\!&
3N H^2 - \frac13 N {\bar R} 
- \frac{N}{3} K^2 - \frac13 N^i K_{,i} 
- \frac{1}{3N} K N^i N^j K_{ij}
\nonumber \\
& \mbox{} &
+ \frac{1}{3N^2} K N^i N^j N_{i ; j}
\;\; . \label{DoPhiIVS3}
\end{eqnarray}

We use our gauge freedom to impose $K \Big\vert_{\rm IVS} = 0$ 
as the gauge condition so that (\ref{DoPhiIVS3}) becomes:
\begin{equation}
\partial_0 {\cal H} \, \Big\vert_{\rm IVS} 
\, = \,
N \Big( 3H^2 - \frac13 {\bar R} \Big)
\;\; . \label{DoPhiIVSK}
\end{equation}
Furthermore, the constraint equation (\ref{constr1}) 
in $K \Big\vert_{\rm IVS} = 0$ gauge is:
\begin{equation}
{\bar R} \, = \, 6H^2 + K_{ij} K^{ij}
\;\; , \label{constr1K}
\end{equation}
implying finally:
\begin{equation}
\partial_0 {\cal H} \, \Big\vert_{\rm IVS} \, = \,
N \Big( H^2 - \frac13 K_{ij} K^{ij} \Big)
\;\; . \label{DoHfinal}
\end{equation}
The lapse function $N$ sets the choice of physical time
as opposed to the coordinate time $t$. Because $K_{ij} 
K^{ij}$ is positive we conclude that the expansion rate 
can indeed diminish. The presence of the diminishing 
term for any value of $H$ indicates that it has the 
ability to completely cancel $H^2$.
\\ [-9pt]

$\bullet \,$ {\it The correspondence limits.} \\
A minimum requirement for our results is to be consistent
with various correspondence limits. Of particular interest 
is the case of de Sitter spacetime. When we consider the 
open coordinate system -- the cosmological patch -- we have:
\begin{equation}
N=1 
\quad , \quad 
N^i = 0 
\quad , \quad
\gamma_{ij} = e^{2Ht} \, \delta_{ij}
\;\; , \label{dSopen1}
\end{equation}
so that -- from (\ref{eom1}) -- we obtain:
\begin{equation}
K_{ij} = -H \gamma_{ij} 
\quad , \quad 
K = -3H
\;\; , \label{dSopen2}
\end{equation}
implying:
\begin{equation}
{\cal H} \, \Big\vert_{\rm IVS} = H
\quad , \quad 
\partial_0 {\cal H} \, \Big\vert_{\rm IVS} = 0
\;\; . \label{dSopen3}
\end{equation}
The expansion rate started at $H$ and stays at $H$.

When we consider the closed coordinate system -- the
full manifold -- we have:
\begin{equation}
N=1 
\quad , \quad 
N^i = 0 
\quad , \quad
\gamma_{ij} = H^{-2} \cosh^2 (H \tau) \, \Omega_{ij}
\;\; , \label{dSclosed1}
\end{equation}
where $\Omega_{ij}$ is the angular line element.
Therefore -- using (\ref{eom1}) -- we get:
\begin{equation}
K_{ij} = - H \tanh(H \tau) \, \gamma_{ij}
\;\; . \label{dSclosed2}
\end{equation}
The choice of $\tau = 0$ as the initial value surface
-- corresponding to the throat of the hyperboloid --
implies that $K_{ij} \, \vert_{\rm IVS} = 0$ and we 
conclude that the system started with no expansion 
and instantaneously began accelerating: 
\begin{equation}
{\cal H} \, \Big\vert_{\rm IVS} = 0
\quad , \quad 
\partial_0 {\cal H} \, \Big\vert_{\rm IVS} = H^2
\;\; . \label{dSclosed3}
\end{equation}

The $\Lambda = 3 H^2 = 0$ limit gives:
\begin{equation}
{\cal H} \, \Big\vert_{\rm IVS} = 0
\quad , \quad 
\partial_0 {\cal H} \, \Big\vert_{\rm IVS} = 
- \frac{N}{3} K_{ij} K^{ij} 
\;\; , \label{LambdaZero}
\end{equation}
leading to contraction when $K_{ij} \neq 0$.

Finally, in the flat spacetime limit:
\begin{equation}
N=1 
\quad , \quad 
N^i = 0 
\quad , \quad
\gamma_{ij} = \delta_{ij}
\;\; , \label{flat}
\end{equation}
the expansion rate ${\cal H}$ vanishes for all time.

\section{Epilogue}
On the initial value surface the expansion rate observable
is proportional to the trace $K$ of the extrinsic curvature,
which is a gauge degree of freedom. It follows that the 
natural gauge choice is $K = 0$ because it allows us to 
start with zero expansion rate and let time evolution 
determine what follows. What we found -- and this is the 
main physical message of our non-perturbative classical 
computation -- is that there exist initial value data 
corresponding to configurations with $K_{ij} \neq 0$ 
which reduce the expansion rate in the presence of a
cosmological constant $\Lambda$. Furthermore, there seems
to be no obstacle for that reduction to completely arrest
the initial expansion due to $\Lambda = 3H^2$. 

Indeed from the constraint equation of motion (\ref{constr2})
and in $K=0$ gauge, the only requirement on the initial
extrinsic curvature is that its covariant divergence vanishes. 
Moreover, the dimensionality of $K_{ij}$ is that of mass and 
there are only two mass scales at our disposal: the inflationary 
scale $H$ and the Planck scale $M_{\rm PL}$. An upper bound 
on $K_{ij} K^{ij}$ cannot vanish with $H$ vanishing because 
there exist configurations with $H = 0 \; \& \; K_{ij} > 0$ 
in direct contradiction. Thus, any upper bound -- if one exists
-- must involve $M_{\rm Pl}$, a situation which still allows 
cancellation of the $H^2$ term in (\ref{DoHfinal}) because 
$M^2_{\rm Pl} \gg H^2$.

It is worth noting that our results confirm again the fact 
that the equation of motion $R = 4 \Lambda$ cannot determine
the physical expansion rate. This was an issue debated in 
\cite{GT} and our present classical calculation -- in which
{\it every} configuration obeys the equation $R = 4 \Lambda$ 
-- shows that {\it any} configuration with $K_{ij} \neq 0$ 
is held together at least infinitesimally and hence deviates
from the de Sitter expansion.

There are many cases in which self-gravitation drastically 
alters the properties of a physical system. For instance, 
it can eliminate the bare mass of a point particle \cite{ADM},
or cause gravitational collapse in a system of incoming
gravity waves \cite{Christodoulou}. The underlying mechanism
can be seen in the Hamiltonian constraint (\ref{constr1}) as 
the interplay between the kinetic energy term -- $K_{ij} 
K^{ij}$ -- and the potential energy term -- the non-linear 
parts buried in $\bar{R}$.

For the physical situation at hand, once inflationary 
gravitons are produced their effect on cosmological
evolution can be understood in completely classical 
terms. Since gravitational waves attract each other and 
act to diminish expansion, when enough of them are present 
they {\it can} completely stop it and even reverse the 
trend leading to collapse. It should be possible to find 
a classical configuration of gravitational waves such that 
the universe holds itself together, against the tendency 
for de Sitter expansion. Such a classical state will almost
certainly {\it not} be completely stable but if it is formed
from the steady production of infrared gravitons over a 
prolonged period of inflation, by causality the decay time 
would almost certainly be {\it longer} than the lifetime 
of the universe.

We do not know what initial value data describe this classical
configuration of gravitons. We do however know from our present 
analysis that initial value data {\it exist} for which the
corresponding configuration does not succumb to accelerated 
expansion. It would be very significant to explicitly verify 
that inflationary graviton production eventually forms a state 
of the kind that stops inflation. Of course even if the latter 
is not completely the case, the very existence of such a 
configuration implies that there is some probability for 
the universe to tunnel to it and, hence, stop inflation.

Finally, we itemize our main conclusions. In the presence of 
a cosmological constant: \\
{\it (i)} The initial value of the expansion rate can be gauged 
to zero; \\
{\it (ii)} The presence of initial gravitational waves with
$K_{ij} \neq 0$ makes the initial time derivative of the 
expansion rate less than its value in de Sitter; \\
{\it (iii)} It seems that nothing precludes initial value data 
which make the initial first derivative of the expansion rate 
vanish; and \\
{\it (iv)} The evolution of the universe is a {\it sustained 
gravitational collapse}.

\vspace{1cm}

\centerline{\bf Acknowledgements}

We should like to thank D. Christodoulou, S. Deser, J. M. Nester, 
and Ch. Soo for discussions. This work was partially supported by the 
European Union (European Social Fund, ESF) and Hellenic national funds 
through the Operational Program ``Education and Lifelong Learning" 
of the National Strategic Reference Framework (NSRF) under the 
``$\Theta\alpha\lambda\acute{\eta}\varsigma$'' action MIS-375734, 
under the ``$A\rho\iota\sigma\tau\epsilon\acute{\iota}\alpha$'' 
action, under the ``Funding of proposals that have received a 
positive evaluation in the 3rd and 4th Call of ERC Grant Schemes''; 
by NSF grant PHY-1205591, and by the Institute for Fundamental Theory 
at the University of Florida.

\end{document}